\documentclass[11pt,twoside]{article}
\usepackage{asp2004}
\usepackage{psfig}
\usepackage{epsf}
\usepackage{graphics}
\usepackage{lscape}
\def\edcomment#1{\iffalse\marginpar{\raggedright\sl#1\/}\else\relax\fi}
\reversemarginpar

\begin{document}
\title{An Infrared Survey of Neutron-Capture Elements in Planetary Nebulae}
\author{N. C. Sterling and Harriet L. Dinerstein}
\affil{University of Texas, Department of Astronomy, 1 University Station, C1400, Austin, TX, 78712-0259}

\begin{abstract}

We present results from an ongoing survey of infrared emission lines from the neutron-capture elements Se and Kr in Galactic planetary nebulae (PNe).  Se and Kr may be produced in the initial steps of the \emph{s}-process during the asymptotic giant branch (AGB) phase of PN progenitor stars, and brought to the surface by convective dredge-up before PN ejection.  Therefore, enriched Se and Kr abundances in PNe indicate that the \emph{s}-process and dredge-up were active within the progenitor stars.  We have detected the emission lines [Kr~III]~2.199 and [Se~IV]~2.287~$\mu$m in 68 of 119 Galactic PNe, and used these line fluxes to derive ionic and total elemental abundances.  Using the ionization correction factors Se$^{3+}$/Se~$\approx$~Ar$^{++}$/Ar and Kr$^{++}$/Kr~$\approx$~S$^{++}$/S, we find a range of Se and Kr abundances, from nearly solar to enriched by a factor of 5-10 times, which implies varying degrees of dredge-up efficiency in the progenitor stars.  We have searched for correlations between \emph{n}-capture element abundances and other nebular properties, and find that PNe with Wolf-Rayet central stars tend to exhibit more elevated Se and Kr abundances than other nebulae.  Bipolar nebulae, believed to arise from the most massive of PN progenitors, may have lower \emph{n}-capture abundances than elliptical PNe.

\end{abstract}
\thispagestyle{plain}

\section{Introduction}

Isotopes of Se and Kr may be produced by slow neutron-capture nucleosynthesis (the ``\emph{s}-process'') during the thermally-pulsing asymptotic giant branch (AGB) phase of planetary nebula (PN) progenitor stars.  Between thermal pulses, free neutrons are released primarily by the reaction $^{13}$C($\alpha, n)^{16}$O in a thin layer between the H- and He-burning shells, then captured by iron-peak seed nuclei.  The seed nuclei undergo successive \emph{n}-captures and $\beta$ decays, and grow progressively heavier (Busso, Gallino, \& Wasserburg 1999).  As the AGB phase proceeds, the thermal pulses are followed by periods of convective dredge-up (third dredge-up, or TDU), in which the convective envelope penetrates the (inactive) H-burning shell, and conveys to the surface material enriched in He-burning and \emph{s}-process products.  The incremental enrichment of the stellar surface with carbon and \emph{s}-process isotopes is the basis of the evolutionary sequence M$\rightarrow$MS$\rightarrow$S$\rightarrow$SC$\rightarrow$C during the AGB phase (e.g.\ Smith \& Lambert 1990).  Through stellar winds and PN ejection, these stars enrich the interstellar medium and contribute to the Galactic chemical evolution of C and \emph{n}-capture elements.

The prevalence of TDU in PN progenitor stars is poorly known, due to the difficulty in deriving carbon abundances in PNe (Kaler 1983; Rola \& Stasi\'{n}ska 1994).  Alternatively, \emph{n}-capture elements can be used to probe the efficiency of TDU in PN progenitor stars, but it was not until 1994 that \emph{n}-capture lines were detected and correctly identified in a PN (P\'{e}quignot \& Baluteau 1994).  In fact, \emph{n}-capture element abundances had been derived for only a handful of PNe before our survey (see also Sterling, Dinerstein, \& Bowers 2002; Sterling \& Dinerstein 2003).  We show that the lines [Kr~III]~2.199 and [Se~IV]~2.287~$\mu$m, identified by Dinerstein (2001), are detectable in a significant fraction of Galactic PNe, and can be used to derive abundances of the \emph{n}-capture elements Se ($Z=34$) and Kr ($Z=36$).  Se and Kr are negligibly depleted onto dust grains in the interstellar medium (Cardelli et al.\ 1993; Cartledge, Meyer, \& Lauroesch 2003), and therefore their gaseous abundances in PNe are expected to represent the total elemental abundances.  These two lines are thus valuable tracers of AGB nucleosynthesis and mixing in PN progenitor stars, and can be used to investigate the yields and Galactic chemical evolution of \emph{s}-process species.

\section{Observations}

We have observed 85 PNe to date in the $K$~band with the CoolSpec spectrometer (Lester et al.\ 2000) on the 2.7-m Harlan J.\ Smith Telescope at McDonald Observatory.  Of these objects, we have detected Se and/or Kr in 53.  Each PN is observed at a resolution $R=500$, which is sufficient to separate [Kr~III]~2.199 and [Se~IV]~2.287~$\mu$m from neighboring features in most nebulae.  The main contaminants to these two lines are H$_2$~3-2~S(3) 2.201 and 3-2~S(2) 2.287~$\mu$m, which are unimportant except in PNe with vibrationally excited H$_2$ (about 15-20\% of our sample).  In PNe exhibiting H$_2$ emission, we use a high resolution setting ($R\approx4500$) to split the [Kr~III]/H$_2$ lines, and thereby estimate the H$_2$ contribution to the unresolvable blend at 2.287~$\mu$m.  In cases where no feature is seen at 2.20~$\mu$m, or if the feature is too weak to detect in the high resolution setting, we assume the maximum contribution of H$_2$ to the 2.287~$\mu$m line, scaled to H$_2$~1-0~S(0) 2.224~$\mu$m (fluorescent excitation; Model~14 of Black \& van~Dishoeck 1987).  We expand our sample by including 34 $K$~band spectra of PNe from the literature (Geballe, Burton, \& Isaacman 1991; Hora, Latter, \& Deutsch 1999; Lumsden, Puxley, \& Hoare 2001), of which 15 exhibit \emph{n}-capture element lines.  Thus, we have found [Kr~III] and/or [Se~IV] in 68 of 119 PNe, including literature data, for an overall detection rate of 57\%.

\section{Abundances and Correlations}

We have derived ionic abundances (or upper limits) for Kr$^{++}$ and Se$^{3+}$ in all of the observed PNe.  We solve for level populations with a five level model atom for Kr$^{++}$ and a two level system for Se$^{3+}$, using transition probabilities from Bi\'{e}mont \& Hansen (1986, 1987) and collision strengths from Sch\"{o}ning (1997).  Note that the [Se~IV]~2.287~$\mu$m collision strength is unknown, so that only relative Se abundances can currently be derived.  We plan to obtain an empirical estimate of this collision strength from observations of Galactic H~II regions, which are not expected to be enriched in \emph{n}-capture elements.

In order to obtain elemental abundances, it is necessary to correct for unseen stages of ionization by using ``ionization correction factors,'' or ICFs.  Based on similar ionization potential ranges, we assume Kr$^{++}$/Kr~$\approx$~S$^{++}$/S and Se$^{3+}$/Se~$\approx$~Ar$^{++}$/Ar, where the S and Ar abundances are generally available in the literature.  We are currently testing the reliability of these ICFs with photoionization models.  The derived Se and Kr abundances exhibit a wide range of values, from near solar (implying little to no TDU) to enriched by a factor of 5-10 times (very efficient TDU).  The average Se and Kr abundances of the full sample are elevated relative to the solar abundance (Table~1), indicating that TDU is prevalent in PN progenitors (but note that we assume the [Se~IV] collision strength is unity; if it is larger, the derived Se abundances would be reduced).

We have explored a number of possible correlations between Se and Kr abundances and other nebular properties, in order to relate the \emph{s}-process and third dredge-up to other nucleosynthetic and morphological traits.  We do not find a correlation between [Kr/H] or [Se/H] and C/O, which is surprising since both \emph{n}-capture elements and C are brought to the stellar surface by TDU.  Previous studies have found a correlation between C/O and \emph{s}-process enhancements in AGB (Smith \& Lambert 1990; Abia et al.\ 2002) and post-AGB stars (Van~Winckel 2003).

\begin{table}[!ht]
\caption{Kr and Se Abundances Vs. Nebular Properties}
\smallskip
\begin{center}
{\small
\begin{tabular}{lcccc}
\tableline
\noalign{\smallskip}
Property & Derived & Number of & Derived & Number of \\
 & $<$Se/H$>^{\rm a}$ & Se Detections & $<$Kr/H$>$ & Kr Detections \\
\noalign{\smallskip}
\tableline
\noalign{\smallskip}
$[$WC$]$ & 3.2$\pm$0.6 (-8) & 13 & 8.7$\pm$3.9 (-9) & 2\\
WELS & 2.2$\pm$0.5 (-8) & 9 & 4.0$\pm$1.6 (-9) & 3\\
Non-[WC]/WELS & 1.8$\pm$0.2 (-8) & 34 & 5.2$\pm$3.5 (-9) & 18\\
\noalign{\smallskip}
\tableline
\noalign{\smallskip}
Type I & 2.3$\pm$0.6 (-8) & 12 & 3.5$\pm$0.9 (-9) & 3\\
Non-Type I & 2.1$\pm$0.2 (-8) & 44 & 5.6$\pm$3.7 (-9) & 20\\
\noalign{\smallskip}
\tableline
\noalign{\smallskip}
Bipolar & 2.3$\pm$0.5 (-8) & 8 & 2.6$\pm$0.9 (-9) & 4\\
Elliptical & 2.9$\pm$0.4 (-8) & 21 & 4.3$\pm$2.1 (-9) & 11\\
Round & 2.4$\pm$0.2 (-8) & 3 & ... & 0\\
Irregular$^{\rm b}$ & 2.0$\pm$0.5 (-8) & 6 & 3.4 (-9) & 1\\
\noalign{\smallskip}
\tableline
\noalign{\smallskip}
Full Sample & 2.2$\pm$0.2 (-8) & 56 & 5.3$\pm$3.4 (-9) & 23\\
Solar$^{\rm c}$ & 2.3$\pm$0.2 (-9) & ... & 1.9$\pm$0.4 (-9) & ...\\
\noalign{\smallskip}
\tableline
\noalign{\smallskip}
\end{tabular}
\begin{itemize}
\item[(a)]{Assumes [Se~IV] collision strength is unity}
\item[(b)]{Includes point-symmetric nebulae}
\item[(c)]{From Lodders 2003}
\end{itemize}
}
\end{center}
\end{table}

In Table~1, we present average Se and Kr abundances for various PN subclasses, along with the mean dispersions and total number of detections for each.  The subclasses we consider include PNe with central stars exhibiting Wolf-Rayet-like features ([WC]) or with weaker and narrower emission lines (weak emission line stars, or WELS), PNe with large N/O and He/H abundances (Peimbert Type~I; Peimbert 1978), and various morphological types.  We find that PNe with H-deficient, C-rich [WC] central stars exhibit larger Se and Kr abundances than other PNe.  This is likely due to the heavy mass-loss and deep mixing these stars underwent as they transformed into [WC] objects (e.g.\ Bl\"{ocker} 2001).    Neutron-capture abundances may be higher in elliptical PNe than bipolar objects, which arise from more massive progenitor stars.  However, the Se abundances of objects classified as Type~I PNe, also believed to derive from massive PN progenitors, are indistinguishable from that of non-Type~I PNe.  We do not find other significant correlations at this stage of our survey.

\section{Conclusions}

The initial results of our survey indicate that [Kr~III]~2.199 and [Se~IV]~2.287~$\mu$m are detectable in a significant fraction of Galactic PNe.  While the derived abundances are somewhat uncertain due to the crude ICFs currently adopted, we find evidence for a range of \emph{s}-process enrichments in PNe, which implies a range of nucleosynthetic yields and dredge-up efficiencies in the progenitor stars.  Our final sample will consist of over 150 PNe, which (with model-derived ICFs) will allow us to more robustly assess the nucleosynthetic histories of PN progenitors and their enrichment of the interstellar medium.

\acknowledgements{This work was supported by NSF grant AST~97-31156.}

\end{document}